\documentclass[aps,pra,floatfix,amsmath,amssymb,twocolumn,showpacs,showkeys]{revtex4}
\usepackage[caption=false]{subfig}
\usepackage{graphicx,bm,color}
\usepackage{amsfonts}

\begin{document}

\title{Entanglement measure using Wigner function: case of generalized vortex state formed by multiphoton subtraction}

\begin{abstract}
The negativity of the Wigner function is discussed as a measure of the non classicality and the quantum interference pattern obtained therein as a possible measure of the entanglement between the two modes of the vortex states. This measure of entanglement is compared with the results obtained from concurrence.
\end{abstract}

\author{Anindya Banerji$^{1}$\footnote{abanerji09@gmail.com}}
\author{Ravindra Pratap Singh$^{2}$\footnote{rpsingh@prl.res.in}}
\author{Abir Bandyopadhyay$^{1}$\footnote{abir@hetc.ac.in}}
\affiliation{$^{1}$Hooghly Engineering and Technology College, Hooghly 712103, India}
\affiliation{$^{2}$Physical Research Laboratory, Ahmedabad 380009, India}

\date{\today}

\pacs{42.65.Lm, 03.65.Ud, 03.67.Mn, 03.67.Bg}
\keywords{Entanglement, Wigner function, Photon addition/subtraction, Quantum optical vortex.}

\maketitle

\section{Introduction}
Over the last decade, the physical characteristics of the entanglement of quantum mechanical states, both pure and mixed, has been recognised as a central resource in various aspects of quantum information processing \cite{Vidal}. Given an arbitrary bipartite pure state $\vert\Psi_{AB}\rangle$, the entropy of entanglement $E(\vert\Psi_{AB}\rangle)$ \cite{Bennett1}, namely the von Neumann entropy of the reduced density matrix $\rho_{A}\equiv\text{Tr}_{B}\vert\Psi_{AB}\rangle\langle\Psi_{AB}\vert$, tells us about the probabilities of transforming $\vert\Psi_{AB}\rangle$ using LOCC, into other pure states in an asymptotic sense. Many efforts have also been devoted to the study of mixed state entanglement \cite{Bennett2, Horodecki}. Given the central status of entanglement, the task of quantifying the degree to which a state is entangled is important for quantum information processing and correspondingly several measures such as entanglement of formation \cite{Bennett2, WoottersPRL}, von Neumann entropy \cite{Nielson} and negativity \cite{Vidal, Zyc} have been formulated. The entanglement of formation or concurrence has been a topic of much interest lately. Although originally formulated to quantify pure state entanglement, it was later extended to mixed states \cite{Munro, Wootters, Wei, Chen}.\\
In this article, we study the entanglement present between the two modes of a bipartite system using the Wigner function. The Wigner function \cite{Wigner} stands out among all (quasi) probability distributions in quantum mechanics \cite{FromOC} because it is real, non – singular, yields correct quantum mechanical operator averages in terms of phase space integrals and possesses definite marginal distributions \cite{Assorted}. We highlight the negativity of the Winger function, which depicts non classicality \cite{Kenfack}, as a measure of entanglement.\\
For our study, we consider optical vortex states formed by quantized radiation fields \cite{GSA97,jbvortex,abir,GSANJP,Jha,anindya}. These are easy to generate non-Gaussian states of light. Optical vortices are phase singularities in the field of light described by a non – separable two dimensional field. It has been shown that such states can be generated from two mode squeezed vacuum under a linear transformation belonging to the SU(2) group with some restrictions \cite{jbvortex}. These can also be generated by coupling squeezed coherent states of two modes using a beam splitter or a dual channel directional coupler \cite{abir}. Recently Agarwal \cite{GSANJP} analyzed quantum optical vortex by subtracting a photon by a 99\% transmitting beam - splitter (BS) from one of the two modes of spontaneous parametric down-conversion (SPDC), generated by the non - linear crystal (NLC). He showed that the subtraction from one mode results into an addition to the other mode; it produces a vortex, which is elliptical in shape.  He also showed that the vortex state had higher entanglement than the two mode squeezed vacuum from which it was generated. It must be noted that a single photon subtraction or addition has already been demonstrated by Parigi et al. \cite{parigi}. A more recent paper studied the enhancement of entanglement, though defined differently, in photon subtraction/addition \cite{CNB}. They studied the enhancement in entanglement generated by subtracting/adding photons in one or both the modes of a two mode squeezed vacuum state. In another work, coherent photon addition/subtraction was considered in which photons were added or subtracted from both the modes simultaneously \cite{new additions}. But the squeezing parameter considered by either is real and hence the resultant states are free of the vortex nature which we consider here. A very interesting feature of the vortex state is the presence of orbital angular momentum (OAM). Vortex states of order \emph{m} carry an OAM of $m\hbar$ per photon. Because of their specific spatial structure and associated OAM, they find various applications in the field of optical manipulation \cite{Grier}, optical communication \cite{Gibson, Rama}, quantum information and computation \cite{Mair}.\\
The paper is organized as follows. In section \ref{vortex}, we introduce the generalized vortex state arising from multi - photon subtraction of the two mode squeezed vacuum. We show that multi - photon subtraction indeed gives rise to vortex states of higher order. We discuss the non classicality of this state with the help of the Wigner function. In section \ref{entanglement}, we discuss the entanglement properties of the generalized vortex state with the logarithmic negativity and entanglement of formation. We also present an interpretation of the quantum interference patterns obtained from the Wigner function as a measure of the entanglement present between the two modes by comparing it with the results obtained from our study of concurrence. We conclude this article after pointing out the significant results and directions for future work in section \ref{conclusion}.

\section{Vortex states arising from photon subtraction/addition}
\label{vortex}
\begin{figure}[!h]
\includegraphics[scale=0.6]{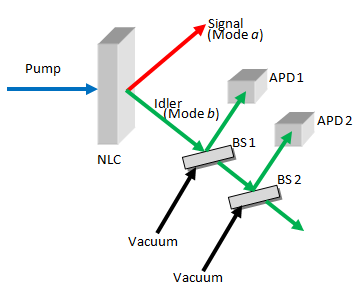}
\caption{(Color online) A schematic of the experimental setup for the production of vortex state of order 2 by subtracting two photons from a two - mode squeezed vacuum using two 99$\%$ beam splitters(BS 1 and BS 2). NLC is a nonlinear crystal. SPDC generates the signal (mode a) and the idler (mode b). APD 1 and 2 are Avalanche Photo Diodes.}
\label{ExpSetup}
\end{figure}

Quantum optical vortices are phase space singularities in the quantized radiation field which possess orbital angular momentum. There are methods of producing these states. One of them is by  mixing two squeezed vacuum modes through BS or dual channel directional coupler \cite{abir}. The other one is by photon subtraction using a BS from one of the outputs of a SPDC \cite{GSANJP,parigi} as shown in Fig. \ref{ExpSetup}. Vortex of higher orders can be produced by increasing the number of BSs and avalanche photo diodes.
Here we discuss the properties of vortices created by subtracting photons.\\
A two-mode squeezed beam is represented by
\begin{equation}
\label{TMS}
\vert \xi \rangle=\text{exp}\left(\xi a^{\dagger}b^{\dagger}-\xi^{*}ab\right)\vert 0,0\rangle, \hspace{0.2 cm} \xi=r\text{e}^{i\phi}
\end{equation}
\noindent where $\xi$ is a complex parameter, \emph{r} is the squeezing parameter and \emph{a} and \emph{b} are the regular bosonic mode operators satisfying the commutation relations $[a, a^{\dagger}]=[b, b^{\dagger}]=1, [a, b^{\dagger}]=0$, etc. The corresponding quadrature distribution is obtained by replacing the bosonic mode operators by the quadrature operators $x, y, p_x$ and $p_y$ which obey the relations $[x, p_x]=[y, p_y]=i$. Here we consider $\hbar=1$. The quadratures $x$, $p_x$ correspond to mode \emph{a} (signal) and $y$, $p_y$ correspond to mode \emph{b} (idler). The quadrature distribution thus obtained has the following form
\begin{eqnarray}
\label{twomodesqueezedvacuum}
\Psi^{(s)}(x, y)=\frac{1}{\sqrt{\left(1-\eta ^2\right)\pi \text{cosh}^2r}}\nonumber\\
\times\text{exp}\left[\frac{2xy\eta-\left(x^2+y^2\right)\eta ^2}{1-\eta ^2}-\frac{1}{2}\left(x^2+y^2\right)\right]
\end{eqnarray}
\begin{figure}[h]
\centering
\subfloat[]{
\includegraphics[scale=0.4]{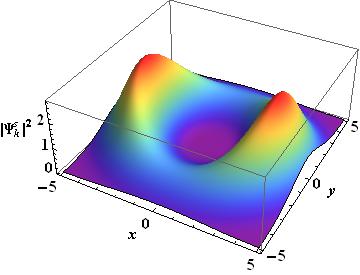}
\label{fig:m3mag}}\\
\subfloat[]{
\includegraphics[scale=0.3]{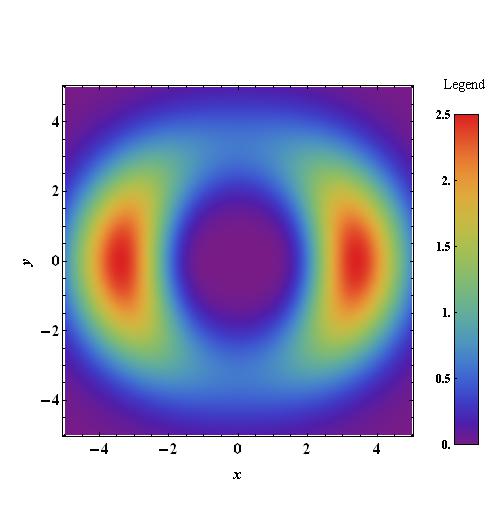}
\label{fig:m3contour}}\\
\subfloat[]{
\includegraphics[scale=0.4]{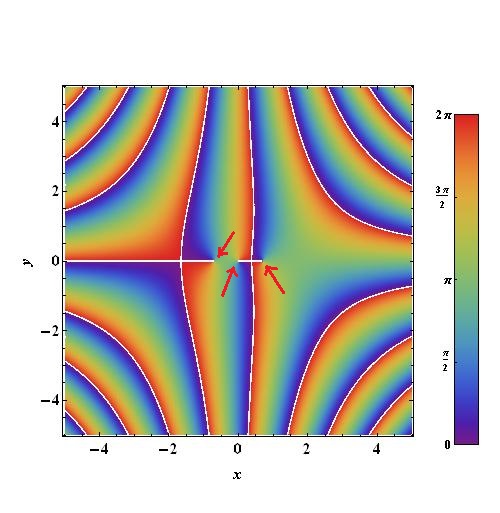}
\label{fig:m3phase}}
\caption{(Color online) Plot of the intensity (a) 3D plot, (b)  contour plot and (c) phase of $\Psi_k^s(x,y)$ for \emph{k}=3. The singularities have been pointed out in Fig. (c).}
\label{fig:m3}
\end{figure}
\noindent The non-separability is evident from the fact that Eq. (\ref{twomodesqueezedvacuum}) cannot be written as a product of two functions each of which are solely dependent on either $x$ or $y$.\\
If an arbitrary number of photons, say \emph{k}, are subtracted from the two mode squeezed vacuum, Eq. (\ref{twomodesqueezedvacuum}) can be generalized to the following \cite{GSANJP}
\begin{eqnarray}
\label{finalstate}
\vert\xi\rangle^{(s)}_k &=&\frac{e^{ik\theta}}{\text{cosh}^{2}r}\sum_{m=0}^{\infty}
e^{im\theta}(\text{tanh}^mr\sqrt{m+k}\vert m+k,m\rangle \nonumber\\
&=& \frac{e^{ik\theta}}{\text{cosh}^2r} a^{\dagger k}\vert\xi\rangle
\end{eqnarray}
We see that the state described in Eq. (\ref{finalstate}) can also be obtained by adding \emph{k} photons to $\vert\xi\rangle$ in the mode \emph{a}. $\vert\xi\rangle^{(s)}_k$ exhibits the property that there is a fixed difference in the number of photons between the signal and idler states. In literature this is one of the so called \textquotedblleft pair coherent states\textquotedblright \cite{GSAPrl, Arvind}. Pair coherent states provide an important example of non classical states of the two mode radiation field. It has been studied in detail for their non classical properties and as examples of EPR states \cite{GSAPrl}.\\
\begin{figure}[h]
\centering
\subfloat[]{
\includegraphics[scale=0.4]{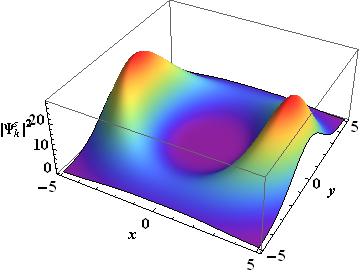}
\label{fig:m4mag}}\\
\subfloat[]{
\includegraphics[scale=0.3]{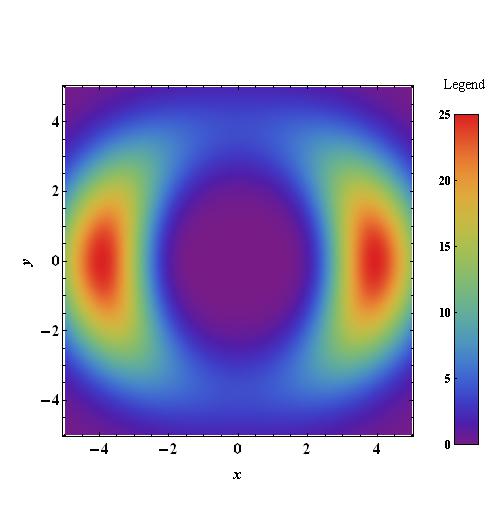}
\label{fig:m4contour}}\\
\subfloat[]{
\includegraphics[scale=0.4]{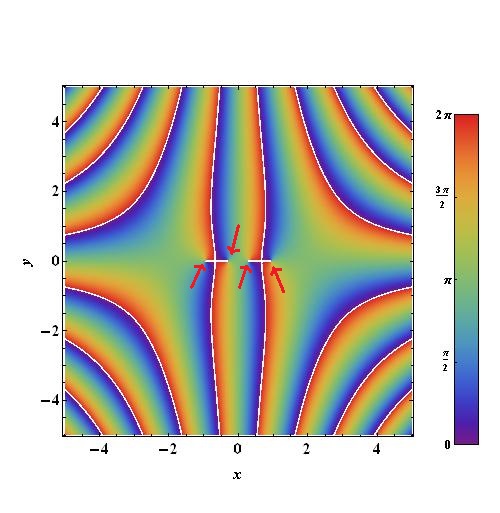}
\label{fig:m4phase}}
\caption{(Color online) Plot of the intensity (a) 3D plot, (b)  contour plot and (c) phase of $\Psi_k^s(x,y)$ for \emph{k}=4. The singularities have been pointed out in Fig. (c).}
\label{fig:m4}
\end{figure}
\begin{figure*}
\centering
\subfloat[$W(x, y)_{p_y=0}^{p_x=0}$]{
\includegraphics[scale=0.3]{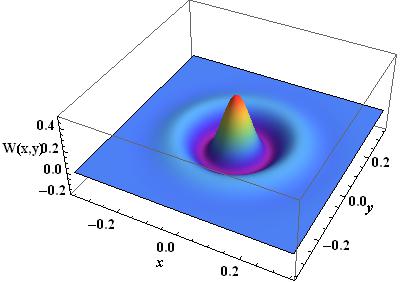}
\label{fig:label-a}}
\qquad
\subfloat[$W(p_x, p_y)_{y=0}^{x=0}$]{
\includegraphics[scale=0.3]{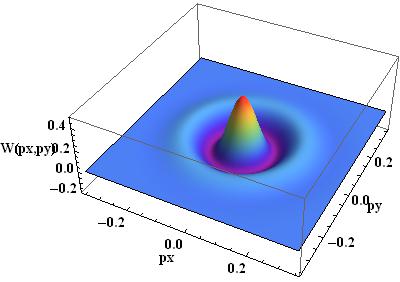}
\label{fig:label-b}}
\qquad
\subfloat[$W(x, p_x)_{p_y=0}^{y=0}$]{
\includegraphics[scale=0.3]{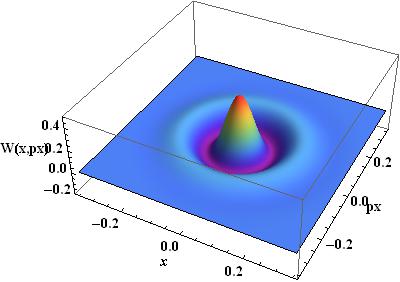}
\label{fig:label-c}}
\qquad
\subfloat[$W(y, p_y)_{p_x=0}^{x=0}$]{
\includegraphics[scale=0.3]{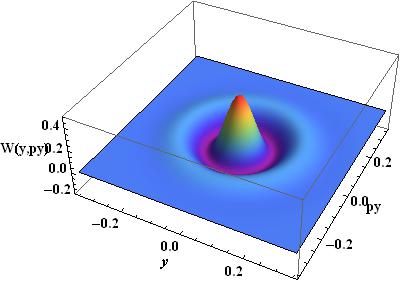}
\label{fig:label-d}}
\qquad
\subfloat[$W(x, p_y)_{y=0}^{p_x=0}$]{
\includegraphics[scale=0.3]{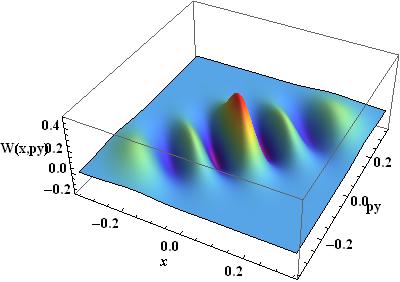}
\label{fig:label-e}}
\qquad
\subfloat[$W(y, p_x)_{p_y=0}^{x=0}$]{
\includegraphics[scale=0.3]{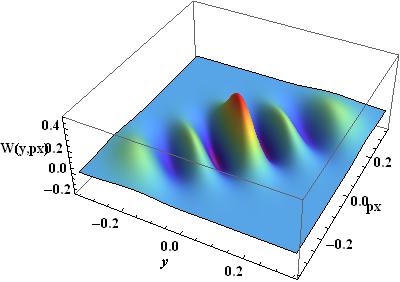}
\label{fig:label-f}}
\caption{(Color online) Wigner function of the TMSV state, for the four photon subtracted case (k=4).}
\label{fig:wigner}
\end{figure*}
The other property of interest is the vortex structure. The quadrature distribution for the state $\vert\xi\rangle^{(s)}_k$,
\begin{equation}
\label{quadDis}
\Psi_k^{(s)}(x,y)=\frac{e^{ik\theta}}{2^{k/2}\text{cosh}r}\left(x-\frac{\partial}{\partial x}\right)^k\Psi^{(s)}(x,y)
\end{equation}
\noindent shows a vortex structure of vorticity \emph{k}. Agarwal \cite{GSANJP} studied the vortex formed for k=1 i.e. by single photon subtraction that can be expressed as
\begin{eqnarray}
\label{quadDisfin}
\Psi_1^{(s)}(x,y)=\frac{\sqrt{2}e^{i\theta}\left(x-\eta y\right)}{\left(1-\eta^2\right)^{3/2}\pi^{1/2}\text{cosh}^2 r}\nonumber \\
\times \text{exp}\left[\frac{2xy \eta-\left(x^2+y^2\right)\eta ^2}{1-\eta ^2}-\frac{1}{2}\left(x^2+y^2\right)\right]
\end{eqnarray}
This is seen to possess a first order vortex structure for $\eta =i|\eta|$. It can be argued that the process of subtracting \emph{k} photons should give rise to vortex states of higher order. As it has already been pointed out, multi-photon subtraction from one of the modes of the two mode squeezed vacuum is similar to adding an equal number of photons in the other mode; the basic idea is to have a difference in the photon number between the two modes. This gives rise to higher order vortex states where the order or vorticity is determined by the difference in number of photons in the two modes. We call these states Two Mode Squeezed Vortex (TMSV) states.\\
\noindent We show the states arising from multiphoton subtraction in Fig. (\ref{fig:m3}) and Fig. (\ref{fig:m4}). It is evident that this gives rise to higher order vortex states. The order can be determined from the phase plots. The singularities in the phase plots have been marked with red arrows (color online). It is observed that the order of the vortex state is equal to the number of photons being subtracted. For example, a vortex state of order 3 is produced when three photons are subtracted and so on. It would thus be interesting to study the entanglement properties of the vortex states with the order of the vortex. In the present article we show usefulness of the Wigner function to study entanglement and nonclassical nature of higher order vortex states formed by quantized radiation field using multiphoton subtraction.\\
\noindent For a number state $\vert n\rangle$, the Wigner function is written as \cite{Walls},
\begin{equation}
W (x, y)=\frac{2}{\pi}\left(-1\right)^n\mathcal{L}_n\left(4q^2\right)e^{-2q^2},
\end{equation}
\noindent where $q^2=x^2 + y^2$ and $\mathcal{L}_n$ is the Laguerre polynomial of order \emph{n}. The Wigner function for a two mode Fock state, $\vert n, m\rangle$, can then be obtained in terms of coherent state parameters $\alpha = x - ip_x$ and $\beta = y - ip_y$ as follows
\begin{eqnarray}
W(\alpha, \beta)&=&\frac{4}{\pi^2}\left(-1\right)^n\mathcal{L}_n
\left[4\vert\alpha\vert^2\right]\exp\left(-2\vert\alpha\vert^2\right)\nonumber \\
&\times &\left(-1\right)^m\mathcal{L}_m\left[4\vert\beta\vert^2\right]\exp\left(-2\vert\beta\vert^2\right)
\end{eqnarray}
The Wigner function for the two mode state $\vert n, 0\rangle$ can then be simplified and written as
\begin{equation}
W(\alpha, \beta)=\frac{4}{\pi^2}\left(-1\right)^n\mathcal{L}_n
\left[4\vert\alpha\vert^2\right]\exp\left[-2\left(\vert\alpha\vert^2 + \vert\beta\vert^2\right)\right].
\end{equation}
It has been shown that if two density matrices are related by a squeezing transformation, then the Wigner functions are related by the following transformation \cite{GSANJP}
\begin{equation}
W(\alpha, \beta) = W(\tilde{\alpha}, \tilde{\beta}),
\end{equation}
\noindent where,
\begin{equation}
\begin{pmatrix} \tilde{\alpha} \\ \tilde{\beta}^*\end{pmatrix} =
\begin{pmatrix}\text{cosh}r && -\text{sinh}r~e^{i\theta}\\ -\text{sinh}r~e^{i\theta} && \text{cosh}r \end{pmatrix} 
\begin{pmatrix} \alpha \\ \beta^* \end{pmatrix}
\end{equation}
\noindent So, the Wigner function for the TMSV, $\vert\xi\rangle_k^{(s)}$, can be written as,
\begin{equation}
W(\tilde{\alpha}, \tilde{\beta})=\frac{4}{\pi^2}\left(-1\right)^k\mathcal{L}_k
\left[4\vert\tilde{\alpha}\vert^2\right]\exp\left[-2\left(\vert\tilde{\alpha}\vert^2 + \vert\tilde{\beta}\vert^2\right)\right],
\end{equation}
\noindent where \emph{k} is the number of photons subtracted. In Fig. (\ref{fig:wigner}) we study the variation of the Wigner function as a function of two of the quadratures out of four (two quadratures for each of the two modes), keeping the other two quadratures as zero.\\
\noindent The non classicality of the TMSV states is evident from the nature of the Wigner function which shows negative regions for all the photon subtracted states. In Fig. (\ref{fig:label-a}) and Fig. (\ref{fig:label-b}), we study the cross correlation between the same quadratures of two different modes. It is observed, though not shown in the figure, that the peak alternates between 0.4 (for even \emph{k}) and -0.4 (for odd \emph{k}). It attains the positive maximum when an even number of photons are subtracted. Negative maximum is attained when an odd number of photons are subtracted. The volume of the negative region of the Wigner functions can be further studied as a measure of the non classicality of these states \cite{Kenfack}. It can also be interpreted as a measure of entanglement between the two modes as explained in section \ref{Concurrence}. Fig. (\ref{fig:label-c}) and Fig. (\ref{fig:label-d}) represents the cross correlation between two quadratures of the same mode. These are exactly similar in nature to that of the cross correlation between same quadratures of different modes. In Fig. (\ref{fig:label-e}) and Fig. (\ref{fig:label-f}), we study the cross correlation between different quadratures of two different modes. These show interesting quantum interference effects. Similar interference pattern has been reported in \cite{jbvortex,Ananth}. We present an interpretation of this quantum inteference later in the article.
\section{Entanglement properties}
\label{entanglement}
In this section we study the mixedness of the TMSV states as well as the entanglement properties of them. 
\subsection{Logarithmic negativity}
There exists various measures of entanglement. A much used tool is the log negativity parameter \cite{Vidal}. We use this to quantatively study the entanglement in the vortex state. It is defined by
\begin{equation}
\label{lognegativity}
\epsilon=\log_2\left(1 + 2\mathcal{N}\right)
\end{equation}
\noindent where $\mathcal{N}$ is the modulus of the sum of all the negative eigenvalues associated with the partial transpose $\rho^{\text{PT}}$ of the density matrix $\vert\xi\rangle_k^{(s)}~^{(s)}_k\langle\xi\vert$. The density matrix of the TMSV state can be written from Eq. (\ref{finalstate}) as follows
\begin{equation}
\label{densitymat}
\rho = \sum_{m,n}c_mc_n\vert m+k, m\rangle\langle n+k, n\vert e^{i\left(m-n\right)\theta}
\end{equation}
\noindent where,
\begin{equation}
c_p = \text{tanh}^pr\frac{\sqrt{p+k}}{\text{cosh}^2r}.
\label{coeff}
\end{equation}
\noindent Here $p = m,~n$. Therefore, the log negativity parameter of Eq. (\ref{lognegativity}) becomes
\begin{equation}
\label{lognegativity2}
\epsilon_k = \log_2\left(\sum_n c_n^2\right)^2.
\end{equation}
\begin{figure}[!h]
\centering
\includegraphics[scale=0.6]{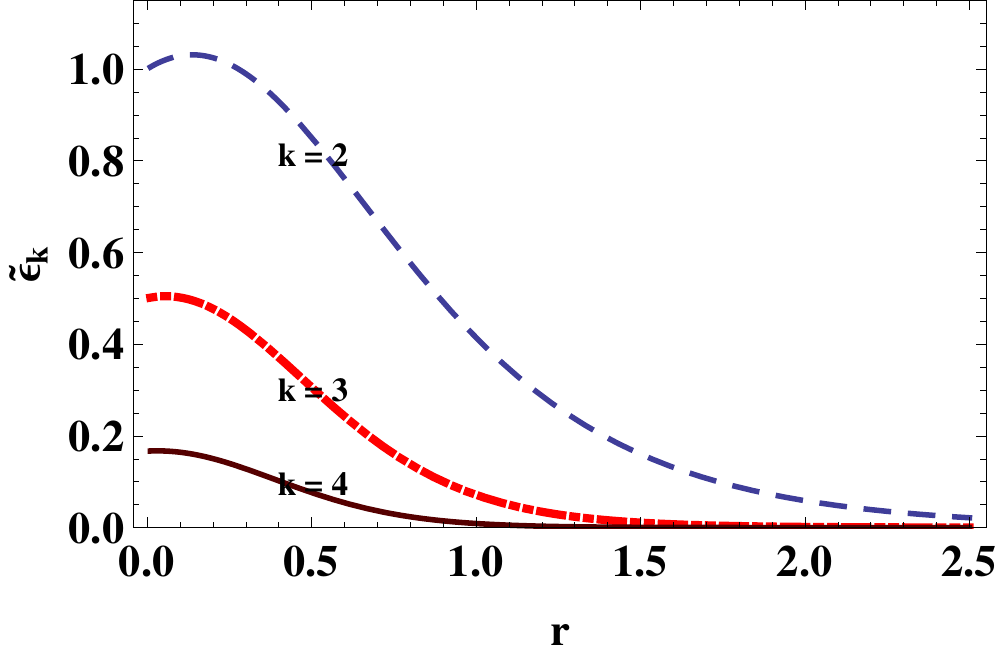}
\label{fig:m3ent}
\caption{(Color online) Entanglement ratio $\tilde{\epsilon_k}$ for the TMSV states}
\label{fig:entanglement-ratio}
\end{figure}
\noindent A similar calculation for the two mode squeezed vacuum gives the result $\log_2\left(e^{2r}\right)$ using $c_n = \text{tanh}^nr/\text{cosh}r$. The entanglement in the vortex state can be compared with that in the state $\vert \xi \rangle$ by studying the ratio of log negativities of the two states. The ratio has the form
\begin{equation}
\label{ratio}
\tilde{\epsilon}_k=\left(\sum_n c_n e^{-r}\right)^2.
\end{equation}
We study the results in Fig. (\ref{fig:entanglement-ratio}). It was observed that entanglement in the TMSV state is reduced with the number of photons subtracted. For two photon subtraction there was a very small rise in entanglement which falls sharply after reaching the peak. When three photons are subtracted the initial value falls to 0.5. For four photon subtraction, the starting value is around 0.17.  This might seem pretty interesting given the nature of the higher order vortex states in Fig. (\ref{fig:m3}) and Fig. (\ref{fig:m4}). When three photons are subtracted, it gives rise to a vortex state of order 3 as can be seen from Fig. (\ref{fig:m3phase}) and vortex of order 4  when four photons are subtracted (Fig. (\ref{fig:m4phase}). This shows that a vortex state of higher order is less entangled than a vortex state of lower order which is counter intuitive. We further investigate the entanglement of the TMSV state in the next section.\\
\subsection{Entanglement of formation}
\label{Concurrence}
In this section we study the entanglement of the TMSV state with the help of concurrence \cite{WoottersPRL,Wootters}, also expressed as the entanglement of formation \cite{Bennett2}. It is a measure of the amount of entanglement needed to create the entangled state. It is defined as \cite{Wei}
\begin{equation}
\label{conc}
E_f=\text{min}_{{p_i,\psi_i}}\sum_i p_iE\left(\vert\psi_i\rangle\langle\psi_i\vert\right)
\end{equation}
\begin{figure}[h]
\includegraphics[scale=0.6]{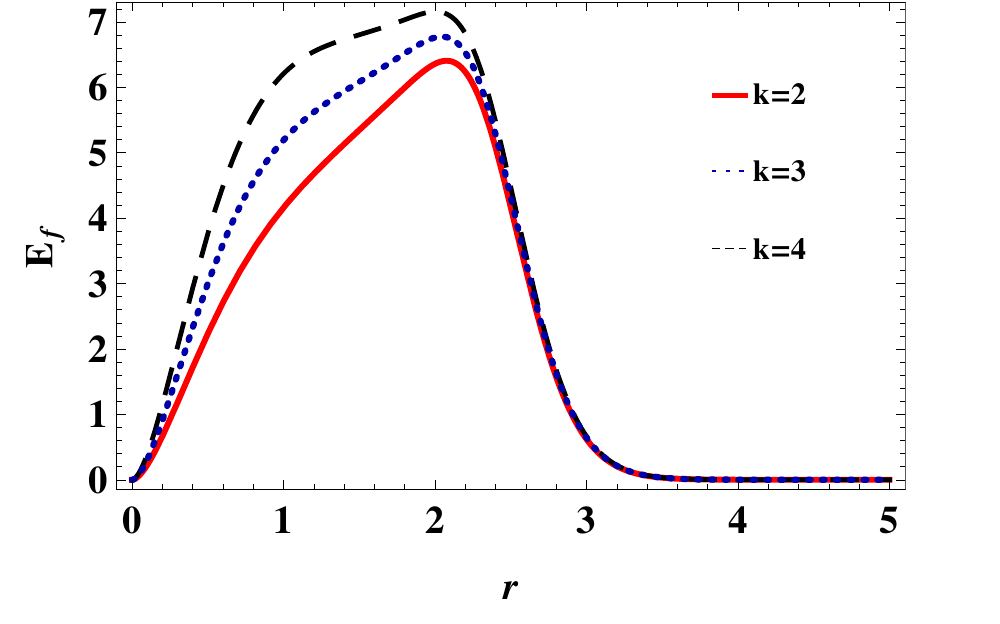}
\caption{(Color Online) Entanglement of formation of the photon added (subtracted) vortex state. \emph{k} denotes the number of photons added (subtracted) in one of the modes.}
\label{fig:conc}
\end{figure}
\noindent The minimization is taken over all possible decompositions of pure states {$\psi_i$} with probabilities $p_i$, which taken together reproduce the density matrix $\rho$ of the given mixed state. On the other hand, $E\left(\vert\psi_i\rangle\langle\psi_i\vert\right)$ is the entropy of entanglement, commonly known as the von Neumann entropy. 
Using Eq. (\ref{densitymat}), we arrive at the following simplified form of the entanglement of formation
\begin{equation}
\label{entformation}
E_f=\text{min}_{c_n,\vert\xi\rangle^{(s)}_k}\sum_{n,m}c_n c_m\left(-\sum_p c_p^2\log_2c_p^2\right)
\end{equation}
\noindent where, $c_i$'s are the same as in Eq. (\ref{coeff}). We study the variation of the entanglement of formation with the squeezing parameter and the order of the vortex in Fig. (\ref{fig:conc}).
\begin{figure}[!h]
\centering
\subfloat[r=1.5]{
\includegraphics[scale=0.3]{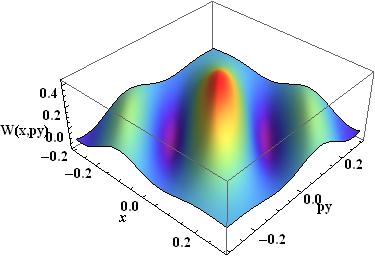}
\label{fig:qi1}}
\qquad
\subfloat[r=2.1]{
\includegraphics[scale=0.3]{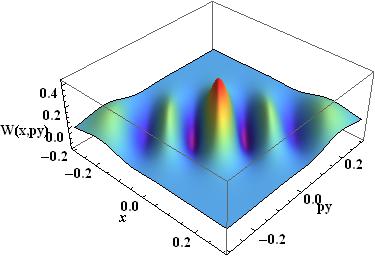}
\label{fig:qi2.1}}
\qquad
\subfloat[r=2.5]{
\includegraphics[scale=0.3]{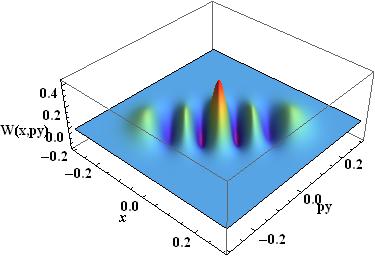}
\label{fig:qi2.5}}
\qquad\\
\subfloat[r=3.5]{
\includegraphics[scale=0.3]{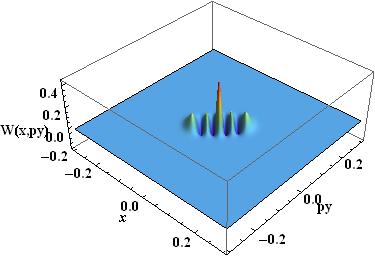}
\label{fig:qi3}}
\caption{(Color online) Wigner function $W(x,p_y)$ of the TMSV states showing quantum interference for different values of \emph{r}, the squeezing parameter, with $k=4$.}
\label{fig:interference}
\end{figure}
We observe that the $E_f$ increases with increase in \emph{r}, the squeezing parameter. The rate of increase gradually slows down till it reaches the maximum value and then falls of rapidly to 0. It should be noticed that $E_f$ attains its maximum value at $r \sim 2.1$. This means that the two modes become more and more entangled with increasing \emph{r} until it reaches a maximum value beyond which the modes start becoming disentangled. We infer that there exists an optimum value of the squeezing parameter, corresponding to which the photon subtracted states exhibit maximum entanglement between the two modes, for a given order. If we go on squeezing the modes more and more, they will no more be entangled. As we move to higher orders of the vortex, the range of values of \emph{r} which produces maximal entanglement, starts broadening. At \emph{k}=4, there exists a well defined range of \emph{r} that produces maximal entanglement of the two modes. The entanglement falls off rapidly on both the ends of this range.\\
We present here an interpretation of the quantum interference effects observed in section \ref{vortex}. In Fig. (\ref{fig:interference}), we study the effect of the squeezing parameter on quantum interference. It is observed that the interference is most prominent for $r=2.1$ (see Fig. \ref{fig:qi2.1}). This observation has a striking similarity with the results obtained from our study of the entanglement of formation. The interference effects decrease as we move away from $r=2.1$ on either side which is exactly similar to what we obtained from the entanglement of formation. Broadening of the range of \emph{r} which gives rise to a well defined interference fringe system with increase in the order of the vortex was also observed. Hence, we suggest that the quantum interefence effects arising from the Wigner function of the cross correlation between different quadratures of the two modes can be interpreted as a signature of the entanglement present between the two modes. It is seen that the number of fringes as well as the volume of the fringes change with changing \emph{r}. However, the maximum number of negative fringes is restricted by the order of the vortex. Since the negativity of the Wigner function is interpreted as a measure of non classicality, we suggest, the volume of the negative part of the quantum interference fringes might be a possible candidate for quantifying entanglement and needs to be further studied.
\section{Conclusion} 
\label{conclusion}
In conclusion, we have studied entanglement between the two modes of the TMSV state using the Wigner function and compared our results using more popular measures of entanglement like logarithmic negativity and entanglement of formation. We observed that the quantum interference effects obtained from the study of the cross correlation between different quadratures of two different modes can be interpreted as arising due to entanglement between the two modes. The volume of the negative region of the interference fringes showed similar variation with the order of the vortex and the squeezing parameter as entanglement of formation.\\
We infer from our study of the entanglement of formation and the quantum interference fringes of the Wigner function, that there exists an optimum value of the squeezing parameter which results in maximum entanglement. This value of the squeezing parameter is independent of the order of the vortex. Our study further points to the need of a more robust measure of entanglement. The TMSV state discussed in this article should also be further investigated to see how it behaves under teleportation.
\acknowledgments{
This work is partially supported by DST through SERB grant no.: SR/S2/LOP - 0002/2011. A Bandyopadhyay acknowledges the Associateship at Physical Research Laboratory, Ahmedabad, where part of the work is done, and discussions with G. S. Agarwal.}

\end{document}